\pdfoutput=1 
\documentclass{JINSTy}

\usepackage[capitalise]{cleveref}
\usepackage{siunitx}
\DeclareSIUnit{\neq}{\ensuremath{\text{n}_\text{eq}}}

\usepackage{xspace}
\usepackage{ifthen}

\newboolean{pdflatex}
\setboolean{pdflatex}{true}

\newboolean{articletitles}
\setboolean{articletitles}{true}

\newboolean{uprightparticles}
\setboolean{uprightparticles}{false}

\newboolean{inbibliography}
\setboolean{inbibliography}{false}

\ifthenelse{\boolean{uprightparticles}}%
{

 \def\PDelta      {\ensuremath{\Delta}\xspace}                 
 \def\PXi      {\ensuremath{\Xi}\xspace}                 
 \def\PLambda      {\ensuremath{\Lambda}\xspace}                 
 \def\PSigma      {\ensuremath{\Sigma}\xspace}                 
 \def\POmega      {\ensuremath{\Omega}\xspace}                 
 \def\PUpsilon      {\ensuremath{\Upsilon}\xspace}                 
 
 \def\PB      {\ensuremath{\mathrm{B}}\xspace}                 
                  
 \def\PD      {\ensuremath{\mathrm{D}}\xspace}

 \def\PK      {\ensuremath{\mathrm{K}}\xspace}

 \def\Pi      {\ensuremath{\mathrm{i}}\xspace}

}
{

 \mathchardef\PDelta="7101
 \mathchardef\PXi="7104
 \mathchardef\PLambda="7103
 \mathchardef\PSigma="7106
 \mathchardef\POmega="710A
 \mathchardef\PUpsilon="7107
                  
 \def\PB      {\ensuremath{B}\xspace}                 
                  
 \def\PD      {\ensuremath{D}\xspace}

 \def\PK      {\ensuremath{K}\xspace}

 \def\Pi      {\ensuremath{i}\xspace}

}

  \def\Kbar  {\kern 0.2em\overline{\kern -0.2em \PK}{}\xspace}

  \def\Dbar    {\kern 0.2em\overline{\kern -0.2em \PD}{}\xspace}

\def\B       {\ensuremath{\PB}\xspace}
\def\Bbar    {\ensuremath{\kern 0.18em\overline{\kern -0.18em \PB}{}}\xspace}

  \def\Y#1S{\ensuremath{\PUpsilon{(#1S)}}\xspace}

\def\Lbar {\ensuremath{\kern 0.1em\overline{\kern -0.1em\PLambda}}\xspace}

\def\AT#1     {\ensuremath{A_{\mathrm{T}}^{#1}}\xspace}           

\def\C#1      {\ensuremath{\mathcal{C}_{#1}}\xspace}                       
\def\Cp#1     {\ensuremath{\mathcal{C}_{#1}^{'}}\xspace}                    
\def\Ceff#1   {\ensuremath{\mathcal{C}_{#1}^{\mathrm{(eff)}}}\xspace}        
\def\Cpeff#1  {\ensuremath{\mathcal{C}_{#1}^{'\mathrm{(eff)}}}\xspace}       
\def\Ope#1    {\ensuremath{\mathcal{O}_{#1}}\xspace}                       
\def\Opep#1   {\ensuremath{\mathcal{O}_{#1}^{'}}\xspace}                    

\newcommand{\tev}{\ifthenelse{\boolean{inbibliography}}{\ensuremath{~T\kern -0.05em eV}\xspace}{\ensuremath{\mathrm{\,Te\kern -0.1em V}}\xspace}}
\newcommand{\gev}{\ensuremath{\mathrm{\,Ge\kern -0.1em V}}\xspace}
\newcommand{\mev}{\ensuremath{\mathrm{\,Me\kern -0.1em V}}\xspace}
\newcommand{\kev}{\ensuremath{\mathrm{\,ke\kern -0.1em V}}\xspace}
\newcommand{\ev}{\ensuremath{\mathrm{\,e\kern -0.1em V}}\xspace}
\newcommand{\gevc}{\ensuremath{{\mathrm{\,Ge\kern -0.1em V\!/}c}}\xspace}
\newcommand{\mevc}{\ensuremath{{\mathrm{\,Me\kern -0.1em V\!/}c}}\xspace}
\newcommand{\gevcc}{\ensuremath{{\mathrm{\,Ge\kern -0.1em V\!/}c^2}}\xspace}
\newcommand{\gevgevcccc}{\ensuremath{{\mathrm{\,Ge\kern -0.1em V^2\!/}c^4}}\xspace}
\newcommand{\mevcc}{\ensuremath{{\mathrm{\,Me\kern -0.1em V\!/}c^2}}\xspace}

\def\barn{\ensuremath{\rm \,b}\xspace}

\def\gsim{{~\raise.15em\hbox{$>$}\kern-.85em
          \lower.35em\hbox{$\sim$}~}\xspace}
\def\lsim{{~\raise.15em\hbox{$<$}\kern-.85em
          \lower.35em\hbox{$\sim$}~}\xspace}

\def\pt         {\mbox{$p_{\rm T}$}\xspace}

\def\tell1  {TELL1\xspace}
\def\ukl1   {UKL1\xspace}

\graphicspath{{figures}{figures/talk}}

\usetikzlibrary{arrows,positioning}

\title{The upgrade of the LHCb Vertex Locator}

\author{T. Bird$^a$ (on behalf of the LHCb VELO group)\\
\llap{$^a$}School of Physics and Astronomy, University of Manchester, Manchester, United Kingdom\\
E-mail: \email{thomas.bird@cern.ch}}

\abstract{
The LHCb experiment is set for a significant upgrade, which will be ready for Run~3 of the LHC in 2020.
This upgrade will allow LHCb to run at a significantly higher instantaneous luminosity and collect an integrated luminosity of \SI{50}{\per\femto\barn} by the end of Run~4.
In this process the Vertex Locator (VELO) detector will be upgraded to a pixel-based silicon detector.
The upgraded VELO will improve upon the current detector by being closer to the beams and having lower material modules with microchannel cooling and a thinner RF-foil.
Simulations have shown that it will maintain its excellent performance, even after the radiation damage caused by collecting an integrated luminosity of \SI{50}{\per\femto\barn}.
}

\keywords{Performance of High Energy Physics Detectors; Particle tracking detectors (Solid-state detectors); Radiation-hard detectors}

\begin{document}


%

\section{The vertex locator}
\label{Section:velo}

\begin{figure}[tbp]
\centering
\includegraphics[height=4.5cm,keepaspectratio]{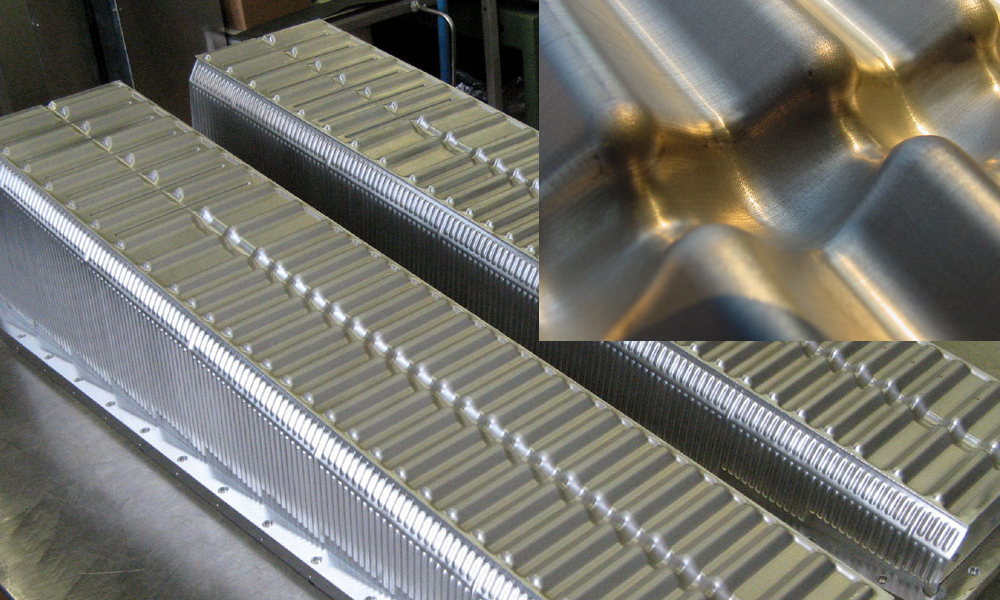}
\hspace{1em}
\includegraphics[height=4.5cm,keepaspectratio]{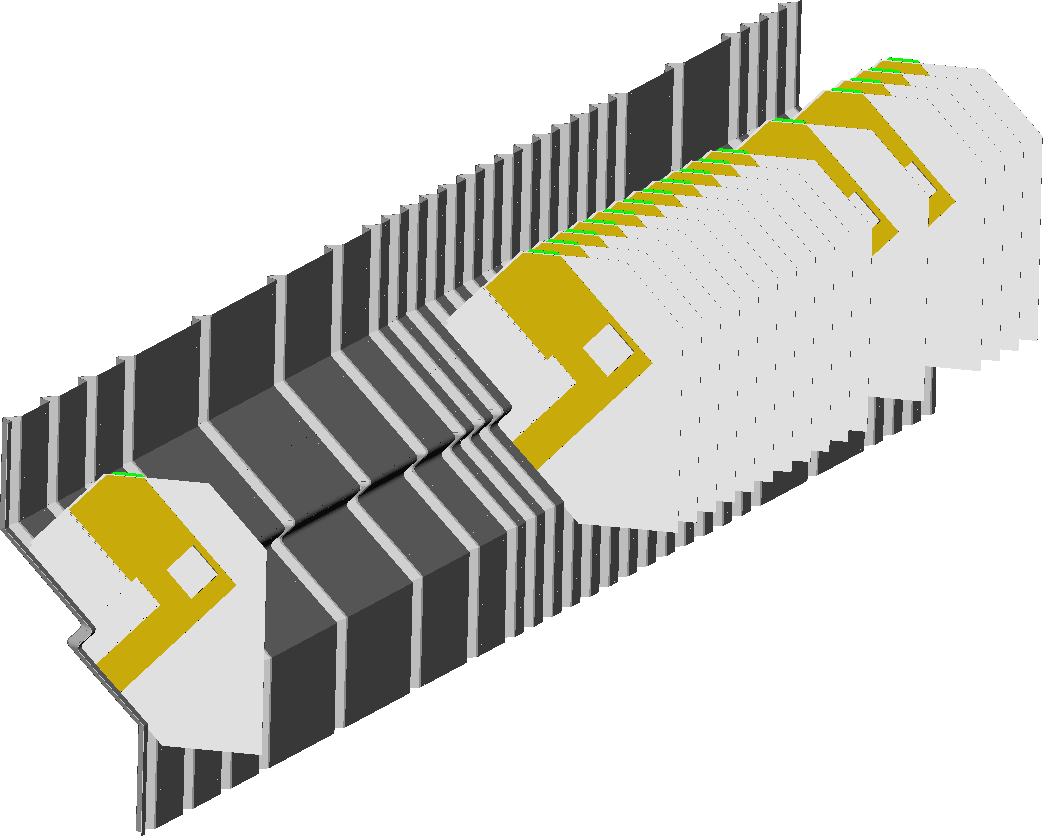}
\caption{
Comparison of the current RF-foil (left) and the upgrade RF-foil (right).
The left image shows the two halves of the current RF-foil, while the inset image zooms into the region of the RF-foil near the beams.
The right image shows one half of the simulated upgrade RF-foil with some modules in place.
}
\label{Figure:rffoil}
\label{Figure:upgraderffoil}
\end{figure}

The LHCb experiment is a forward single-arm spectrometer collecting data at the Large Hadron Collider (LHC) at CERN.
The Vertex Locator (VELO) detector allows LHCb to measure vertices very accurately, such that primary vertices (from proton-proton collisions) and secondary vertices (from decays of short-lived particles) can  be separated.
The VELO is a silicon strip detector made of 42 modules, where each module has strips in the $R$ and $\phi$-directions on alternating sides.
The modules are placed along the beam line over roughly a meter, such that each particle which originates from the interaction region at an angle in the range from \SIrange{15}{390}{\milli\radian} can be reconstructed.
The VELO modules are cooled using evaporative $\text{CO}_2$ cooling.
The modules are kept in a vacuum, which is separated from the LHC vacuum by a \SI{300}{\micro\meter} thick RF-foil~\cite{reoptimisedtdr}; the foil stops the beams from inducing currents in the modules.
\Cref{Figure:rffoil} shows the complex shape of the RF-foil.
Nominally the VELO is \SI{27}{\milli\meter} from the LHC beams to ensure they do not damage the VELO.
During data taking, once the beams are stable, the modules move in so that the sensor edges are \SI{8.2}{\milli\meter} from the beams.
More details about the VELO and LHCb can be found in~\cite{reoptimisedtdr,velotdr,lhcbdetector}.
The performance of the VELO and the effects of radiation damage are discussed in~\cite{performance,radiation}.

\section{The LHCb upgrade}
\label{Section:lhcbupgrade}

The LHC does not deliver the maximum instantaneous luminosity that it could to LHCb; the beams are separated to reduce the number of collisions.
The general purpose detectors (ATLAS and CMS) have a maximum instantaneous luminosity of \SI{7e33}{\per\square\centi\meter\per\second}, whereas LHCb receives \SI{4e32}{\per\square\centi\meter\per\second}, almost a factor of 20 less.
The hardware trigger in LHCb uses transverse energy ($E_t$) in the calorimeters to trigger events with interesting hadronic decays.
When the luminosity is increased a harder cut must be made on $E_t$ to keep data rates manageable, but this reduces the trigger efficiency.
This means the trigger yield saturates as luminosity is increased~\cite{lhcbupgradeloi}.
To overcome this limitation LHCb is planning an upgrade during LHC long shut-down 2 (LS2).
This upgrade will allow the full detector to be read out into a software trigger, bypassing the hardware trigger; this will enable LHCb to run at a luminosity of \SI{2e33}{\per\square\centi\meter\per\second}.
This rise in luminosity increases the occupancy of the VELO to unacceptable levels.
Further, running in these conditions pushes the radiation damage beyond the maximum allowable and so the VELO will need to be upgraded.

\section{The vertex locator upgrade}
\label{Section:upgrade}

The upgrade VELO will use pixel sensors~\cite{veloupgradetdr} instead of strip sensors; this is the largest change between the current and the upgrade VELO.
Aside from this, many aspects will remain the same: the modules will move in close to the beams when they become stable, evaporative $\text{CO}_2$ cooling will be used and most of the support infrastructure outside of the vacuum will remain the same.
In the next section the upgraded detector will be described and the major differences highlighted.

\subsection{Detector overview}
\label{Section:upgrade:improvements}

\begin{figure}[tbp]
\centering
\includegraphics[width=0.8\textwidth,keepaspectratio]{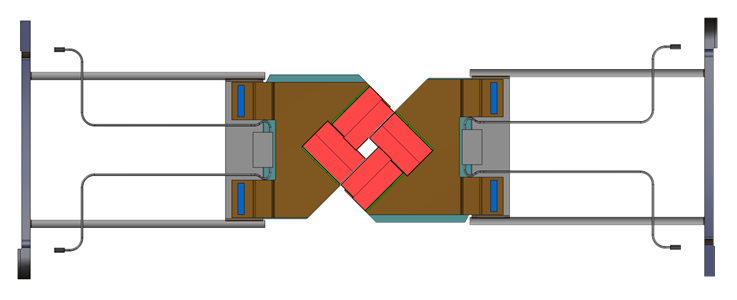}
\caption{
The diagram shows the upgrade VELO modules and one possible design of the supports which will hold the modules in place.
The silicon substrate (turquoise) with microchannel cooling inside is shown, however it is mostly hidden beneath the hybrid (brown).
The sensors (red) will be evenly distributed on opposite sides of the module.}
\label{Figure:modules}
\end{figure}

Each VELO upgrade module will contain 4 planar silicon sensors, as can be seen in \cref{Figure:modules}.
Each sensor will be read out by 3 VeloPix chips, which are based on Timepix3~\cite{timepix}.
The VeloPix measures $14.07\times{}14.07\,$\si{\square\milli\meter} and contains $256\times{}256$ pixels, which leads to a pitch of \SI{55}{\micro\meter}.
The VeloPix will use a binary data-driven readout of $4\times{}2$ super-pixels.
The sensors will need a bias voltage of \SI{1000}{\volt} after the radiation damage received while collecting an integrated luminosity of \SI{50}{\per\femto\barn}~\cite{veloupgradetdr}.
It is difficult to have this high voltage without sparking between the sensor and the VeloPix and so a guard ring of \SI{450}{\micro\meter} has been specifically designed to reduce the risk of sparking.
Additionally the sensor edge may be coated to further reduce the risk of sparking.
The VeloPix and sensors will be bump-bonded and together glued onto the silicon substrate, which will be cooled using microchannels.
These are made by etching \SI{200}{\micro\meter} wide and \SI{120}{\micro\meter} deep channels into \SI{260}{\micro\meter} thick silicon.
Then another \SI{140}{\micro\meter} thick piece of silicon is overlaid to seal the channels.
Liquid $\text{CO}_2$ is pumped through these channels at a pressure of \SIrange{20}{30}{\bar}~\cite{veloupgradetdr} where it evaporates and cools the module.
Unlike other methods of cooling, there will be no mismatch in coefficients of thermal expansion and so the module will not deform when cooled and heated.
Further, this is a particularly low-material cooling method since it requires no additional conductive material around the sensors.
Thermal simulations of the microchannel cooling have determined that the tip of the sensor could extend \SI{5}{\milli\meter} from the edge of the substrate, which further reduces the material that particles must pass through.
The microchannels must not leak into the VELO vacuum; to ensure this, the pressure was successfully cycled a thousand times from \SIrange{0}{160}{\bar} using a prototype with no leakage.
The active silicon sensor will be moved almost $40\%$ closer to the beams, from \SI{8.2}{\milli\meter} to \SI{5.1}{\milli\meter}; this reduces the extrapolation distance to the vertices and so improves the vertex and IP resolution.
The RF-foil is responsible for $50\%$ of material that particles pass through in the VELO, so its design is critical.
The RF-foil is shaped to fit tightly around the VELO upgrade modules and is corrugated to minimise its contribution to the material; this leads to a very complicated shape as pictured in \cref{Figure:upgraderffoil}.
It will be milled out of a solid aluminium block to a thickness of \SI{250}{\micro\meter} and then possibly chemically thinned around the area near the beams.

\subsection{Simulated performance}
\label{Section:upgrade:performance}

\begin{figure}[tbp]
\centering
\includegraphics[width=0.32\textwidth,keepaspectratio]{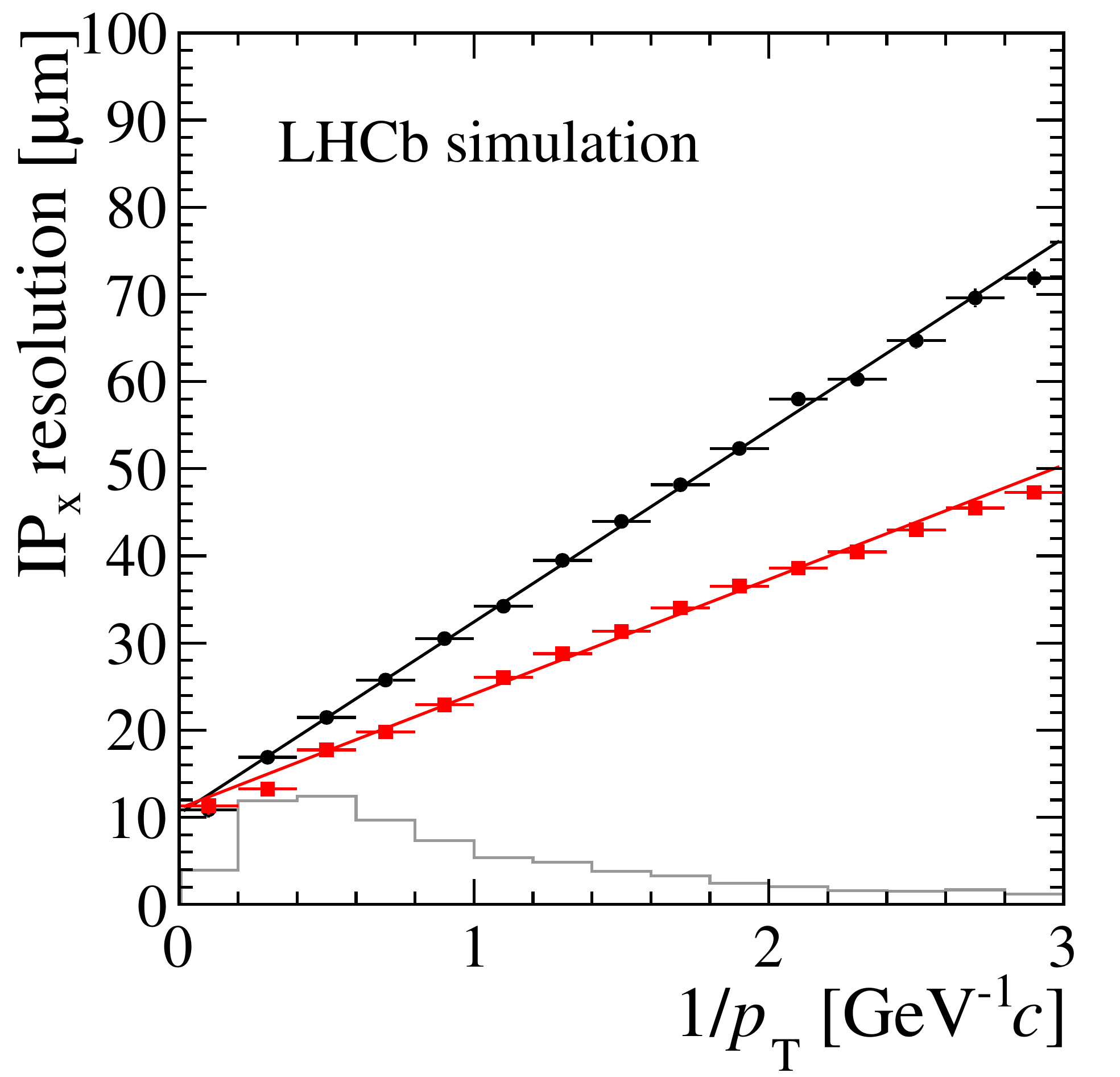}
\includegraphics[width=0.32\textwidth,keepaspectratio]{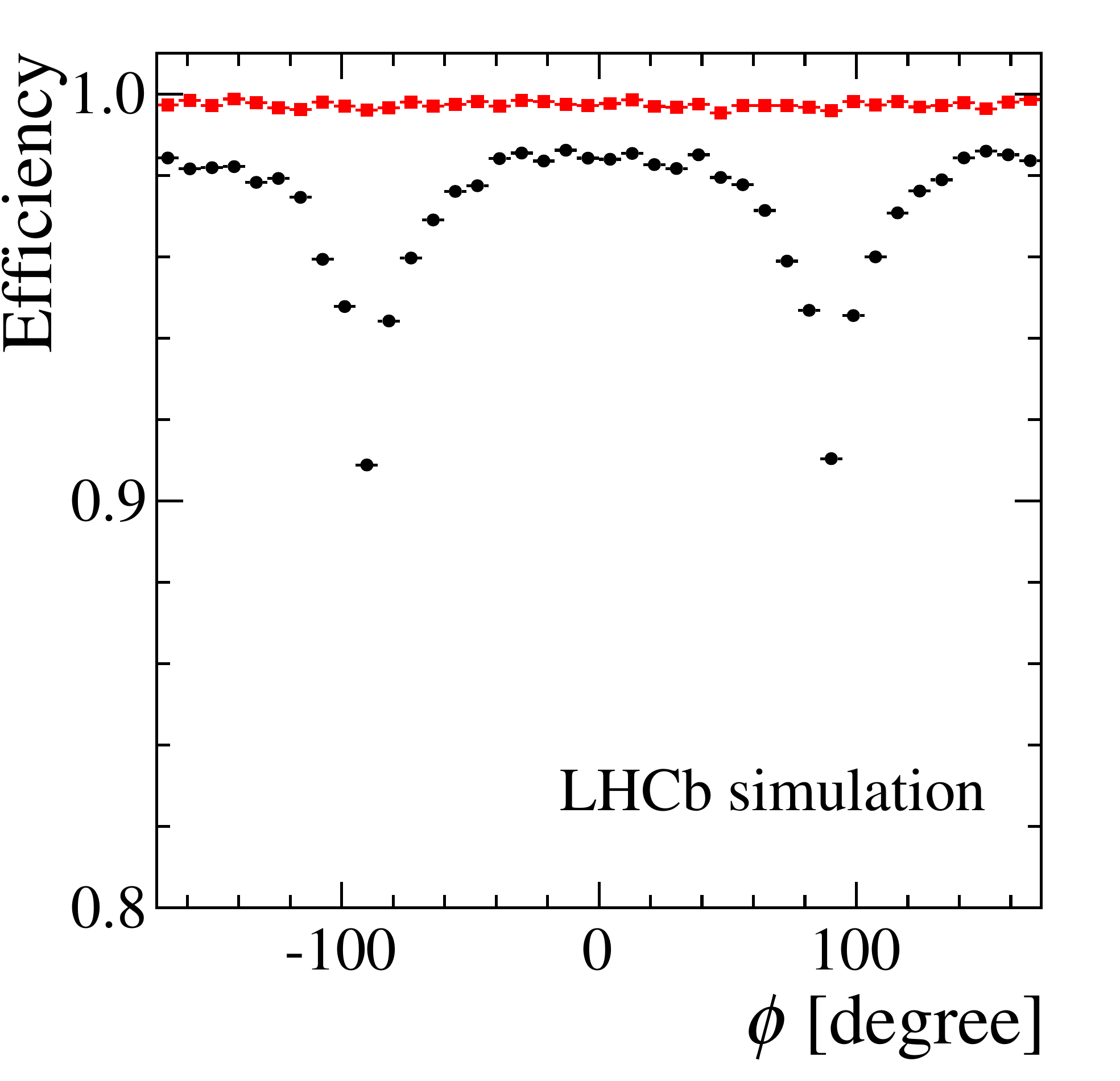}
\includegraphics[width=0.32\textwidth,keepaspectratio]{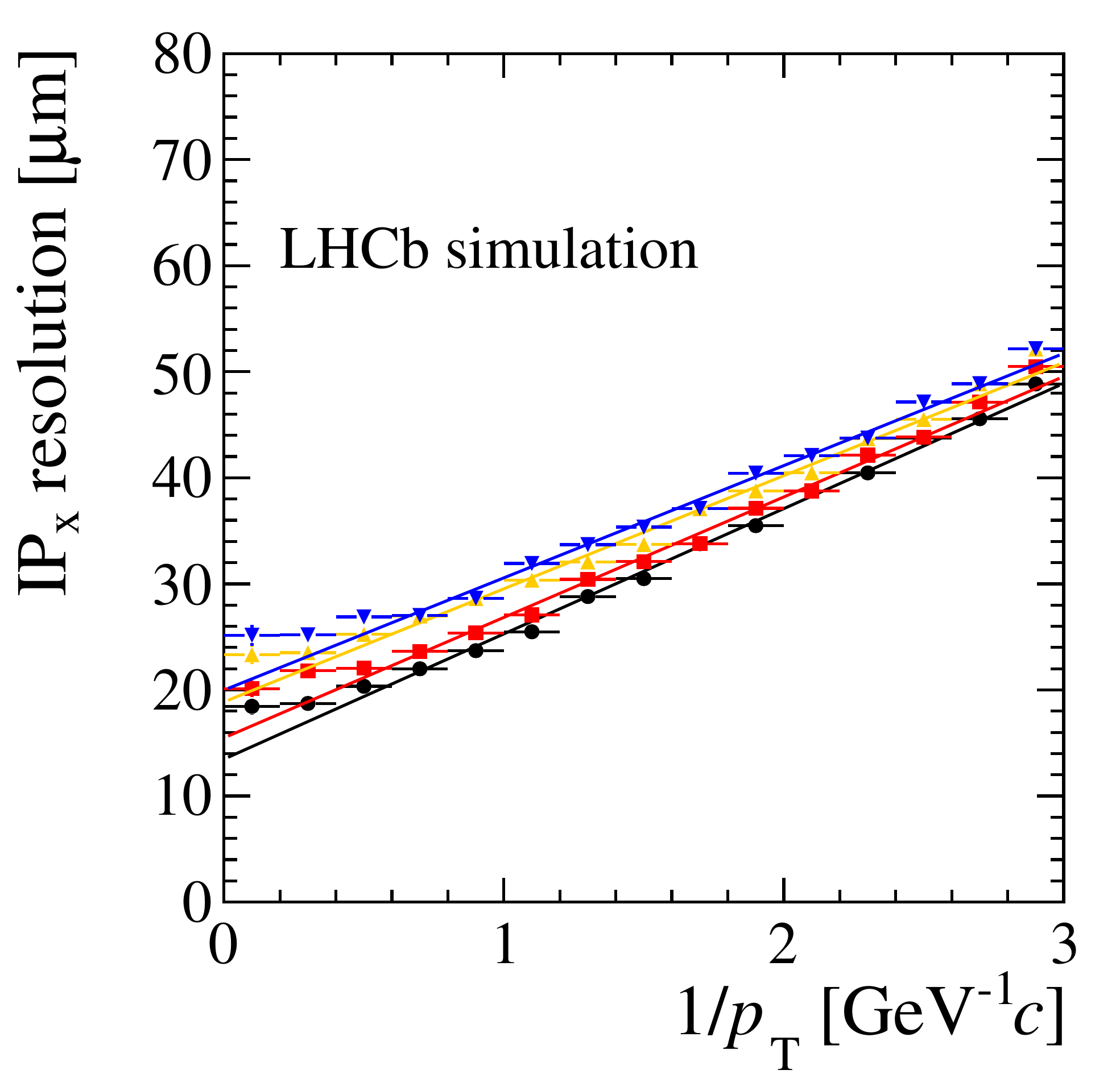}
\caption{
The left plot shows the IP resolution as a function of $1/\pt{}$ for the current VELO (black circles) and the upgrade VELO (red squares), with a typical $1/\pt{}$ spectrum in grey.
The middle plot shows tracking efficiency as a function of azimuthal angle ($\phi$) for the current and upgrade VELO detectors.
The right plot shows the effect of radiation damage on the IP resolution; the points show the performance after \SI{0}{\per\femto\barn} (black circles), \SI{10}{\per\femto\barn} (red squares), \SI{30}{\per\femto\barn} (yellow upward triangles) and \SI{50}{\per\femto\barn} (blue downward triangles).
}
\label{Figure:ipres}
\label{Figure:trackingefficiency}
\label{Figure:radiationdamage}
\end{figure}

The impact parameter (IP) of a track is the perpendicular distance between the track direction and its associated primary vertex; this is an important property for LHCb, because it effects physics quantities such as the lifetime resolution and the IP can also be used to trigger events which contain long-lived particles like \B{} mesons.
In \cref{Figure:ipres} the IP resolution is shown as a function of inverse transverse momentum ($1/\pt{}$), where it can be seen that the upgrade VELO will significantly improve the IP resolution.
The improvements come from the low-material modules, the thinner RF-foil and from moving the active silicon closer to the beams.
The expected tracking efficiency of the current and upgrade VELO detectors during Run~3 is shown in \cref{Figure:trackingefficiency}.
The increase in occupancy in Run~3 creates many fake tracks in the current strip-based VELO and reduces its tracking efficiency, but the upgrade pixel-based VELO does not suffer from this issue.
The excellent performance of the upgrade VELO is only useful if it can withstand the effects of being irradiated.
\Cref{Figure:radiationdamage} shows the expected effect of radiation on the performance of the upgrade VELO with sensors biased to \SI{500}{\volt}.
There is only a relatively small degradation in performance after collecting an integrated luminosity of \SI{50}{\per\femto\barn}, which can almost completely be removed by further increasing the sensor bias.
This demonstrates that the detector can maintain the excellent performance even after heavy irradiation.

\section{Conclusion}
\label{Section:conclusion}

The LHCb experiment is set for a significant upgrade, which will be ready for Run~3 of the LHC in 2020.
This upgrade will allow LHCb to run at a significantly higher instantaneous luminosity and collect an integrated luminosity of \SI{50}{\per\femto\barn} by the end of Run~4.
In this process the VELO will be upgraded to a pixel-based detector.
The upgraded VELO will improve upon the current detector by being closer to the beams and having lower material modules with microchannel cooling and a thinner RF-foil.
Simulations have shown that it will maintain its excellent performance, even after the radiation damage caused by collecting an integrated luminosity of \SI{50}{\per\femto\barn}.




%

\end{document}